\documentclass[prb,showpacs,twocolumn,longbibliography]{revtex4-1}

\usepackage[english]{babel}
\usepackage[utf8]{inputenc}
\usepackage{amsmath}
\usepackage{amssymb}
\usepackage[caption=false]{subfig}
\usepackage{amssymb}
\usepackage{epsfig}
\usepackage{graphicx}
\usepackage{amsmath}
\usepackage{array,color}
\usepackage{natbib}

\usepackage[usenames,dvipsnames]{xcolor}
\definecolor{forestgreen}{rgb}{0.11,0.54,0.15}
\definecolor{purple}{rgb}{0.62,0.10,0.96}
\definecolor{dockerblue}{rgb}{0.11,0.56,0.98}
\definecolor{freeblue}{rgb}{0.25,0.41,0.88}

\usepackage[pdftex,plainpages=false,colorlinks=true,linkcolor=Red, citecolor=blue, urlcolor=blue]{hyperref}

\makeatletter
\renewcommand{\p@subsection}{}
\renewcommand{\p@subsubsection}{}
\makeatother
\begin{document}

\title{Antagonistic effects of nearest-neighbor repulsion on the superconducting pairing dynamics in the doped Mott insulator regime}
\author{A. Reymbaut$^1$, M. Charlebois$^1$, M. Fellous Asiani$^1$, L. Fratino$^2$, P. S\'emon$^1$, G. Sordi$^2$, and A.-M. S. Tremblay$^{1,3}$}
\affiliation{
$^1$D\'{e}partement de physique and Institut quantique, Universit\'{e} de Sherbrooke, Sherbrooke, Qu\'{e}bec, Canada J1K 2R1 \\
$^2$Department of Physics, Royal Holloway, University of London, Egham, Surrey, UK, TW20 0EX\\
$^3$Canadian Institute for Advanced Research, Toronto, Ontario, Canada, M5G 1Z8
}
\date{\today}
\begin{abstract}
The nearest-neighbor superexchange-mediated mechanism for $d_{x^2-y^2}$ superconductivity in the one-band Hubbard model faces the challenge that nearest-neighbor Coulomb repulsion can be larger than superexchange. To answer this question, we use cellular dynamical mean-field theory (CDMFT) with a continuous-time quantum Monte Carlo solver to determine the superconducting phase diagram as a function of temperature and doping for on-site repulsion $U=9t$ and nearest-neighbor repulsion $V=0,2t,4t$.  In the underdoped regime, $V$ increases the CDMFT superconducting transition temperature $T_c^d$ even though it decreases the superconducting order parameter at low temperature for all dopings. However, in the overdoped regime $V$ decreases $T_c^d$. We gain insight into these paradoxical results through a detailed study of the frequency dependence of the anomalous spectral function, extracted at finite temperature via the MaxEntAux method for analytic continuation. A systematic study of dynamical positive and negative contributions to pairing reveals that even though $V$ has a high-frequency depairing contribution, it also has a low frequency pairing contribution since it can reinforce superexchange through $J=4t^2/(U-V)$. Retardation is thus crucial to understand pairing in doped Mott insulators, as suggested by previous zero-temperature studies. We also comment on the tendency to charge order for large $V$ and on the persistence of $d$-wave superconductivity over extended-$s$ or $s+d$-wave.
\end{abstract}
\maketitle



\section{Introduction}

In BCS theory,\cite{BCS:1957_1,BCS:1957_2} the exchange of virtual phonons mediates an attraction between electrons at low frequencies while the direct Coulomb repulsion acts over a much larger energy scale. This allows the repulsive component of this interaction to be screened out through the  Anderson-Morel mechanism,~\cite{Morel:1962,Scalapino:1966} which leads to the pseudopotential $\mu^*$ in the more refined Migdal-Eliashberg theory. \cite{Migdal:1958,Eliashberg:1960} In essence then, retardation is key to the mechanism of conventional BCS superconductivity. 

Cuprate superconductors, that we study here, are found in the vicinity of antiferromagnetic or Mott insulating states. This demonstrates the presence of sizable on-site repulsion between electrons, excluding conventional BCS $s$-wave pairing. But even before the discovery of cuprates, it was suggested, based on extensions of arguments by Kohn and Luttinger,\cite{kohn:1965} that the exchange of antiferromagnetic fluctuations would lead to $d$-wave superconductivity in the vicinity of an itinerant antiferromagnet.\cite{Beal-Monod:1986,Scalapino:1986} Many different methods\cite{Bickers:1987, Bickers_dwave:1989, Moriya:1990, Monthoux:1991, ScalapinoThread:2010, ZanchiSchulz:1998, ZanchiSchulz:2000, Halboth:2000a, Halboth:2000b, Honerkamp:2001a, Honerkamp:2001, HonerkampSalmhoferTflow:2001, HonerkampSalmhoferRice:2002, Tsai:2001, MetznerReview:2012, Mraz:2003, Kyung:2003, MonthouxScalapino:1994, DahmTewordt:1995, ManskeEremin:2003, AbanovChubukovNorman:2008, Hassan:2008, Raghu:2010} have by now shown that this is a viable mechanism for $d$-wave superconductivity in the presence of repulsion. Similarly, in the strong correlation limit, namely for doped Mott insulators, several approaches\cite{Miyake:1986, Kotliar:1988, Giamarchi:1991, Paramekanti:2001, Paramekanti:2004, AndersonVanilla:2004, LeeRMP:2006, PathakShenoyRanderiaTrivedi:2009} have found $d_{x^2-y^2}$ superconductivity. All of these previous results are based on studies of variants of the Hubbard model, a model that was suggested early on~\cite{Anderson:1987a} as containing the key to cuprate superconductivity.

Generalizations of dynamical mean-field theory,\cite{Georges:1996,Maier:2005,KotliarRMP:2006,LTP:2006} which are particularly suited for the strong correlation limit but are also an excellent guide to the physics at weak to intermediate correlation strength,\cite{Maier:2000a,maier_d:2005,Senechal:2005,AichhornAFSC:2006,Aichhorn:2007,Haule:2007,Kancharla:2008,Kyung:2009,gull_superconductivity_2013,SordiSuperconductivityPseudogap:2012} suggest that pairing is maximized in the intermediate regime, where the on-site interaction $U$ is of order the bandwidth $W=8t$. \footnote{Some non-perturbative calculations based on weak correlation ideas even agree at intermediate coupling~\cite{LTP:2006} with strong-correlation based approaches.}   

In all the above approaches, spin fluctuations with either an antiferromagnetic or a singlet character have been argued to drive the pairing. For strong correlations, the characteristic energy scale of these fluctuations, the exchange interaction $J$, is given by $4t^2/U$ and the $d_{x^2-y^2}$ gap symmetry adopted by the Cooper pairs allows them to avoid the direct effect of the on-site repulsion $U$ because of the node in the two-electron wave function (Pitaevskii-Br\"uckner). Yet, we know that the Coulomb interaction is not perfectly screened and that the effect of the nearest-neighbor repulsion (or extended Hubbard interaction) $V$, for example, cannot be eliminated even in a $d$-wave state. On the contrary, one might consider $V$ to be very detrimental to superconductivity. Roughly, we expect the effective interaction to be the difference $J-V$ so that superconductivity could disappear for $V > J$. From the value of the Coulomb interaction computed  at nearest-neighbor distance with a relative dielectric constant of order 10, we estimate $V\approx 400$ meV while $J$ is measured to be\cite{HaydenJHeisenberg:1991} $J\approx 130$ meV. The presence of highly polarizable charge layers may weaken $V$,\cite{RaghuStrongCoupling:2012,AshcroftTallon:2012} yet this remains an important question of principle. For very small correlations ($U\ll W$), it has been argued~\cite{Raghu:2012} that pairing is destroyed as soon as $V\geq U(U/W)$, close to the FLEX result.~\cite{OnariExtended:2004} None of the other calculations~\cite{Zhou-tJV:2004,NoackWhiteSupra:1995,NoackBulutScalapino:1997,ArrigoniHarjuKivelson:2002,Raghu:2012} suggest that pairing can survive for $V>4J$, except for a variational calculation~\cite{Plekhanov:2003} and a zero-temperature calculation with cellular dynamical mean-field theory.\cite{SenechalResilience:2013}

The effect of $V$ is also important as a matter of principle because, as we saw in the BCS case, its influence is deeply related to the crucial question of retardation that is seldom discussed in the above approaches. Yet, this question remains an unsolved problem even for eminent physicists.\cite{Anderson:2007,Scalapino_E-letter, Scalapino_RMP:2012} The question of the ``glue'',~\cite{Civelli:2009,CivelliPairingPRL:2009,Kyung:2009,SenechalResilience:2013,Gull_Millis:2014} namely the existence or not of retardation, has been addressed also recently in models that include the effect of oxygen.\cite{Varma:2016} 

In this paper, we use plaquette dynamical mean-field theory~\cite{Haule:2007} to extend the zero-temperature study of the effect of $V$ on $d_{x^2-y^2}$ superconductivity performed in reference \onlinecite{SenechalResilience:2013} to finite temperature using a continuous-time quantum Monte Carlo solver. The finite temperature results are not necessarily the same as the zero-temperature ones since it has been shown\cite{Fratino:2016} that $T_c^d$ does not scale like the zero-temperature order parameter, contrary to BCS theory. We define $T_c^d$ as the temperature below which short-ranged $d_{x^2-y^2}$ superconducting pairs begin to form. In two dimensions, a Kosterlitz-Thouless vortex binding transition would occur at a lower temperature. In the presence of small interplane coupling, a three-dimensional transition would occur at a temperature slightly larger than the Kosterlitz-Thouless transition.\cite{Hohenberg:1967, Kosterlitz_Thouless:1973} As in Ref.~\onlinecite{SenechalResilience:2013}, we display antagonistic effects of $V$.

Furthermore, we address the question of retardation, which explains the apparent paradoxical effects of $V$, through a detailed study of the dynamics of pairing contained in the Gorkov function\cite{Gorkov:1958} (also called the anomalous Green's function)
$-\langle \hat{\mathcal{T}}_{\tau}\, \hat{c}_{i\uparrow}(\tau)\, \hat{c}_{j\downarrow}(0) \rangle_{\hat{\mathcal{H}}}$. 
In the same way that the dependence at $\tau=0$ of $\langle \hat{c}_{i\uparrow}\, \hat{c}_{j\downarrow} \rangle_{\hat{\mathcal{H}}}$ on the distance between the sites $i$ and $j$ tells us about the superconducting correlation length, the frequency dependence for two fixed neighboring sites informs us on the pairing dynamics. While direct real-frequency studies are sometimes possible,~\cite{Maier:2008,Kyung:2009,SenechalResilience:2013, Sakai:2014, Sakai:2015} calculations at finite temperature rely on Matsubara-frequency or imaginary-time calculations. For quantum Monte Carlo data in particular, one must use maximum entropy methodology for analytic continuation to the real axis.\cite{Jarrell:1996} However, the maximum entropy analytic continuation of the Gorkov function is usually not trivial, despite the best efforts,~\cite{Gull_Millis:2014, Gull_Millis:2015} because of its sign-changing spectral weight. Nevertheless, the new MaxEntAux method \cite{Reymbaut:2015_MaxEnt}, that we take advantage of here, has recently enabled one to perform maximum entropy analytic continuation of the Gorkov function by using an auxiliary spectral weight.

The model and methods are introduced in section \ref{sec:model}. In particular, we justify the cluster size and the impurity solver that we use. Including the nearest-neighbor repulsion $V$ requires an additional approximation that is also explained. In this section on methods, we then introduce the finite temperature definition of the anomalous spectral weight and corresponding cumulative order parameter. These quantities necessitate the analytic continuation of the Gorkov function to obtain the anomalous spectral weight, a quantity that is odd in frequency and not necessarily positive on the positive real axis. Nevertheless, the maximum entropy method can be used to perform the analytic continuation with the MaxEntAux method that we briefly describe. The results in section~\ref{sec:results} show the antagonistic effects of $V$ and show that this can be understood from a more detailed look at the pairing dynamics. One finds that there is retardation. Pairing is controlled by superexchange whose value is modified by $V$.  There is also a depairing contribution from $V$ that has both retarded and instantaneous pieces. We conclude this section on results by discussing charge fluctuations. Further discussion of the results including additional comments on the glue and pairing mechanism can be found in Sec.~\ref{sec:discussion}.

\section{Model and methods} \label{sec:model}

In addition to the model, we discuss here the extension of the cellular dynamical mean-field method necessary to include the effect of nearest-neighbor repulsion and the extension of the maximum entropy method necessary to obtain the finite-temperature pairing dynamics.

\subsection{Extended Hubbard model}

We study the extended Hubbard model on the square lattice, namely
\begin{align}
\hat{\mathcal{H}} & = -t\sum_{\langle i, j\rangle \; \sigma} \left(\hat{c}^\dagger_{i\sigma} \, \hat{c}_{j\sigma}+ h.c.\right) + U \sum_{i} \hat{n}_{i\uparrow} \, \hat{n}_{i\downarrow} \nonumber \\
 & \qquad \qquad + V \sum_{\langle i, j \rangle} \hat{n}_{i} \, \hat{n}_{j} - \mu \sum_i \hat{n}_i\, ,
\label{Extended_Hubbard_model}
\end{align}
where $t$ is the nearest-neighbor hopping, $U$ is the local part of the Coulomb repulsion, $V$ is the nearest-neighbor Coulomb repulsion, and $\mu$ is the chemical potential  that is set so that the system is hole-doped, although with only nearest-neighbor hopping on the square lattice, electron- and hole-dopings are equivalent. For all numerical results, we work in energy units where $t=$~1.  We consider the on-site interaction strength $U=9t \equiv 9$ and three values for $V$: $V=0$, $V=2$, and $V=4$. The value $U=9$ is larger than $U_{\mathrm{MIT}} \sim 6$ where the Mott transition occurs at half-filling in the approach described below. We are thus in the doped Mott insulator regime, where the effect of $V$ is less important than in the weak correlation case.\cite{SenechalResilience:2013}  

\subsection{Cellular dynamical mean-field theory}\label{sec:model_Cellular}

We work with cellular dynamical mean-field theory (CDMFT) \cite{Kotliar:2001,Maier:2005,KotliarRMP:2006,LTP:2006} where a 2$\times$2 cluster of sites is dynamically coupled to a bath of non-interacting electrons through a frequency-dependent hybridization function that is determined self-consistently. The quantum impurity problem is solved with a continuous-time quantum Monte Carlo (CTQMC) solver \cite{Gull:2011} in the hybridization expansion (CT-HYB),~\cite{werner:2006,WernerMillis:2006,haule:2007i} especially suited in the strong-correlation limit.~\cite{Gull:2007} Other CTQMC solvers based on the weak coupling expansion scale better with system size than CT-HYB and have been used extensively to do larger cluster sizes.~\cite{Gull:2009,Gull_Millis:2012,gull_superconductivity_2013,Gull_Millis:2014,Gull_Millis:2015} However, they have a severe sign problem in the doped Mott insulator regime. That is why we are restricted to the CT-HYB solver and to the 2$\times$2 plaquette for the value $U=9$ that we study. Our version of the code is state of the art. It includes optimization with lazy skip-list~\cite{SemonSkipList:2014} and four point updates that insure ergodicity in the case of superconductivity~\cite{Semon:2014}. 

The advantage of the approach is that dynamical correlations arising from short-range physics are taken into account exactly without mean-field decoupling on the cluster itself. The disadvantage is that long-wavelength fluctuations are taken into account only in a mean-field way through static mean-field order parameters. Here, for example, the CDMFT bath breaks $\mathcal{U}(1)$ symmetry in the superconducting state. Superconducting quantities, such as the superconducting order parameter, are computed using cluster Green's functions. We neglect long-range antiferromagnetic and charge orders.  Normally, we could frustrate antiferromagnetism through a next-nearest-neighbor hopping amplitude $t'$. However, this term worsens the fermionic sign problem, so it is neglected here. Indications of a tendency towards charge ordering are discussed further in subsection \ref{sec:results}\textcolor{Red}{.}\ref{SubSec_CDW}. 

The convergence to the thermodynamic limit of various methods has been benchmarked for a number of methods recently.~\cite{LeBlancBenchmark:2015} Here, we do not aim for quantitative accuracy. Instead, we take the following point of view. Recent work~\cite{ReymbautThese:2016,Reymbaut:2016_Crossovers} on the crossover regime at temperatures above the Mott transition at half-filling, shows that even if the position in the $T-U$ plane of the critical point ending the Mott transition at half-filling depends on frustration and cluster size, the qualitative behavior of the crossovers at sufficiently high temperature is independent of these effects.~\cite{Reymbaut:2016_Crossovers} Long-range antiferromagnetic order can hide some of the crossovers, but not all, and when antiferromagnetism is removed by frustration, the crossovers are revealed. We expect similar behavior in the present case. The normal state is controlled by a finite-doping first-order transition and associated crossover regimes.~\cite{SordiSuperconductivityPseudogap:2012,Sordi:2013} Most of the features of the superconducting dome are controlled by the normal-state properties, including pseudogap, and hence the first-order transition becomes an organizing principle for the superconducting state.~\cite{Fratino:2016}  Our results should thus be similar to those we would obtain in a situation where antiferromagnetism is strongly suppressed by frustration. It is very likely, however, that long-ranged antiferromagnetism would displace the superconducting dome and compete with superconductivity in the strongly underdoped regime.~\cite{Reymbaut:2016_Crossovers} 

\subsection{Including nearest-neighbor repulsion $V$}

The derivation of CDMFT, based for example on the self-energy functional approach,~\cite{Potthoff:2003a,Potthoff:2011} rests on the fact that the interaction is local. Applying CDMFT to the case where the nearest-neighbor repulsion $V$ is present requires a further approximation. Even for clusters, it is possible to use an extended version of dynamical mean-field theory (DMFT) that decomposes the near-neighbor interaction $V$  with Hubbard-Stratonovich fields and treats the resulting fermion-boson theory in the spirit of DMFT.~\cite{Si_EDMFT:2000,Chitra_Kotliar_EDMFT:2000,Pseudogap_t-J_EDMFT_Haule:2003,Haule:2007} The equivalent approach has been used for the $t$-$J$ model. It was found~\cite{Haule:2007} that the results are qualitatively similar, whether DMFT or extended DMFT is used.

It is also possible to use a simpler approximation to include $V$ in CDMFT simulations, the so-called dynamical Hartree approximation.~\cite{AichhornPotthoffExtended:2004,SenechalResilience:2013} In that approach, $V$ is taken into account {\it exactly on the cluster} and in the Hartree approximation between clusters. Since we neglect long-range charge order, the effect of inter-cluster $V$ reduces in that case to a shift in chemical potential. The second figure of Ref.~\onlinecite{SenechalResilience:2013} shows that this approximation can be trusted to extract the physics of antiferromagnetism in the normal state so that no qualitative change should be expected in the present work. One could add the normal and anomalous Fock contractions to the Hartree contractions of the inter-cluster interactions considered here. Investigating these corrections will be the subject of future studies. 

\subsection{Anomalous spectral weight and cumulative order parameter for pairing dynamics}

Consider the Gorkov function describing $d$-wave superconductivity in position space 
\begin{equation}
\mathcal{F}_{ij}(\tau) = -\left\langle \hat{\mathcal{T}}_{\tau}\, \hat{c}_{i\uparrow}(\tau)\, \hat{c}_{j\downarrow}(0) \right\rangle_{\hat{\mathcal{H}}} \, .
\end{equation}
At $\tau=0$, if one computes  $\mathcal{F}_{ij}(0)$  as a function of distance between $i$ and $j$, one finds that this function decays with the superconducting coherence length. Analogously, the real-time dependence of $\mathcal{F}_{ij}$ at fixed distance (nearest-neighbor for $d$-wave and on-site for $s$-wave) gives us information on the characteristic frequencies involved in pairing. More specifically, the frequency information is in the spectral weight of the Gorkov function at nearest-neighbor distance, which we call the anomalous spectral weight $\mathcal{A}^{an}_{ij}(\omega)$. 

Another useful quantity to characterize the pairing dynamics is to study the cumulative order parameter\cite{Kyung:2009} defined by
\begin{equation}
\mathcal{I}_{\mathcal{F}}^{ij}(\omega) = \int_{-\omega}^\omega \! \frac{\mathrm{d}\omega'}{2\pi} \, \mathcal{A}^{an}_{ij}(\omega')\, f(-\omega')
\label{I_F}
\end{equation}
where $f(-\omega')=[1+e^{-\beta\omega'}]^{-1}$ is the Fermi-Dirac distribution. In the limit $\omega \rightarrow +\infty$, the spectral representation shows that this quantity is just the sum over all Matsubara frequencies or equivalently the $\tau=0$ limit of the Gorkov function, the $d$-wave superconducting order parameter 
\begin{equation}
\mathcal{I}^{ij}_{\mathcal{F}}(\omega \to +\infty) = \left\langle  \hat{c}_{i\uparrow}\, \hat{c}_{j\downarrow} \right\rangle_{\hat{\mathcal{H}}}\, ,
\label{I_F_inf_bis}
\end{equation}
taken positive by convention here. At zero temperature, Eq.~\eqref{I_F} reduces to the formulas in Refs.~\onlinecite{Kyung:2009,SenechalResilience:2013}. The cumulative order parameter Eq.~\eqref{I_F} is basically the integral of the anomalous spectral weight over positive frequencies. It converges to the order parameter at large frequencies. Positive contributions to $\mathcal{A}_{ij}^{an}(\omega)$ increase the order parameter, hence are considered pair-forming, whereas negative contributions are considered pair-breaking. 

The BCS and Eliashberg cases have been worked out in Ref.~\onlinecite{Kyung:2009} to illustrate the usefulness of the anomalous spectral weight and of the corresponding cumulative order parameter. 

For a CDMFT solution of the 2$\times$2 plaquette, only the wave-vectors $(\pi,0)$ or $(0,\pi)$ give finite $d$-wave anomalous spectral functions (of opposite signs because of $d$-wave pairing symmetry). Let us then only consider $\vec{k}=(\pi,0)$. Looking at this particular wave-vector is equivalent to considering Cooper pairs made from two electrons on nearest-neighbor sites $i,j$. Indeed, take the Fourier transform $\mathcal{F}_{\vec{k}}$. Numbering the 2$\times$2 cluster sites from 1 to 4 clockwise, one has
\begin{eqnarray}
\mathcal{F}_{(\pi,0)} & = & \mathcal{F}_{11} - \mathcal{F}_{12} + \mathcal{F}_{14} - \mathcal{F}_{13} \\
\mathcal{F}_{(0,\pi)} & = & \mathcal{F}_{11} + \mathcal{F}_{12} - \mathcal{F}_{14} - \mathcal{F}_{13}.
\end{eqnarray}
Since there is a node along the diagonal for $d$-wave pairing, $\mathcal{F}_{11}=\mathcal{F}_{13}=0$  and $\mathcal{F}_{12} = -\mathcal{F}_{14}$.  This gives $\mathcal{F}_{(\pi,0)} = -\mathcal{F}_{(0,\pi)}= 2\mathcal{F}_{14} = -2\mathcal{F}_{12}$. Hence, we define the $d$-wave superconducting order parameter $\varphi_{SC}$ calculated on the plaquette as 
\begin{eqnarray}
\mathcal{I}_{\mathcal{F}}(\omega \to +\infty) & = & 2\,  \mathcal{I}^{ij}_{\mathcal{F}}(\omega \to +\infty) \nonumber \\ 
 & = & \left\langle  \hat{c}_{\vec{k}\uparrow}\, \hat{c}_{-\vec{k}\downarrow} \right\rangle_{\hat{\mathcal{H}}}^{\vec{k}=(\pi,0)} \nonumber \\
 & = & \varphi_{SC}\, .
\label{I_F_inf}
\end{eqnarray}

\subsection{MaxEntAux method for analytic continuation}

The previous discussion shows that we need the spectral weight for the Gorkov function, 
\begin{equation}
\mathcal{F}(\vec{k},i\omega_n)= - \int_0^\beta  \! \mathrm{d}\tau \, e^{i\omega_n \tau} \left\langle \hat{\mathcal{T}}_{\tau}\, \hat{c}_{\vec{k}\uparrow}(\tau)\, \hat{c}_{-\vec{k}\downarrow}(0) \right\rangle_{\hat{\mathcal{H}}}\, .
\end{equation}
This does not have a positive spectral weight and hence cannot be trivially analytically continued by maximum entropy methods. The MaxEntAux method~\cite{Reymbaut:2015_MaxEnt} has been developed recently to address this problem. That method consists in defining an auxiliary Green's function (built in order to have a positive spectral weight) that expands as a sum of normal Green's functions (which have positive spectral weights) and anomalous Green's functions.~\cite{Reymbaut:2015_MaxEnt} The auxiliary Green's function is defined as
\begin{equation}
\mathcal{G}_{aux}(\vec{k},\tau) = -\left\langle \hat{\mathcal{T}}_\tau\, \hat{a}_{\vec{k}}(\tau)\, \hat{a}_{\vec{k}}^\dagger(0) \right\rangle_{\hat{\mathcal{H}}}
\label{G_aux}
\end{equation}
where
\begin{equation}
\hat{a}_{\vec{k}} = \hat{c}_{\vec{k}\uparrow} + \hat{c}^\dagger_{-\vec{k}\downarrow}\, .
\end{equation}
Expanding the product of operators in Eq.~\eqref{G_aux}, moving to Matsubara frequency space and using inversion symmetry, one finds
\begin{equation}
\mathcal{G}_{aux}(\vec{k},i\omega_n) = \mathcal{G}_{\uparrow}(\vec{k},i\omega_n) - \mathcal{G}_{\downarrow}(\vec{k},-i\omega_n) + 2\mathcal{F}(\vec{k},i\omega_n)
\end{equation}
where
$\mathcal{G}_{\sigma}(\vec{k},i\omega_n) = -\int_0^\beta \! \mathrm{d}\tau \, e^{i\omega_n \tau}\langle \hat{\mathcal{T}}_\tau\, \hat{c}_{\vec{k}\sigma}(\tau)\, \hat{c}_{\vec{k}\sigma}^\dagger(0) \rangle_{\hat{\mathcal{H}}}$ denotes the normal Green's functions for spin $\sigma$. When there is no broken time-reversal symmetry, one can finally extract the anomalous spectral function $\mathcal{A}_{an}(\vec{k},\omega)$, containing all relevant frequencies for pairing, via
\begin{equation}
\mathcal{A}_{an}(\vec{k},\omega) = \frac{1}{2} \left[ \mathcal{A}_{aux}(\vec{k},\omega) - \mathcal{A}(\vec{k},\omega) - \mathcal{A}(\vec{k},-\omega) \right]
\end{equation}
where $\mathcal{A}_{\uparrow}(\vec{k},\omega) = \mathcal{A}_{\downarrow}(\vec{k},\omega) \equiv \mathcal{A}(\vec{k},\omega)$. In the case of singlet $d$-wave superconductivity (with no breaking of time-reversal symmetry), the anomalous spectral function is real and odd in frequency. The analytic continuation giving $\mathcal{A}_{aux}(\vec{k},\omega)$ and $\mathcal{A}(\vec{k},\omega)$ can be carried out with any Maximum Entropy analytic continuation code, but we find the OmegaMaxEnt program~\cite{Bergeron:2015} most robust and useful. In particular, its diagnostic tools allow a check on the accuracy of the analytic continuation. 


\section{Results}\label{sec:results}

Paradoxically, $V$ seems to both enhance and decrease superconductivity, depending on doping and temperature range,~\cite{SenechalResilience:2013} a phenomenon discussed in section~\ref{sec:antagonistic}. Section \ref{sec:retardation} introduces the first evidence of retardation in the pairing mechanism and the role of superexchange $J$. The pair-forming mechanism coming from $V$ is consistent with a magnetic pairing mechanism, as explained in section~\ref{sec:magnetic}. The pair-breaking effect of $V$ is partly instantaneous, as revealed by the frequency dependence of the anomalous self-energy shown in section \ref{sec:instantaneous}. When $V$ is too large, there is a clear tendency towards charge ordering, as discussed in section~\ref{sec:charge}.

\subsection{Antagonistic effects of $V$ on the superconducting order parameter} \label{sec:antagonistic}

\begin{figure}[h!]
	\begin{center}
		\includegraphics[width=0.49\textwidth]{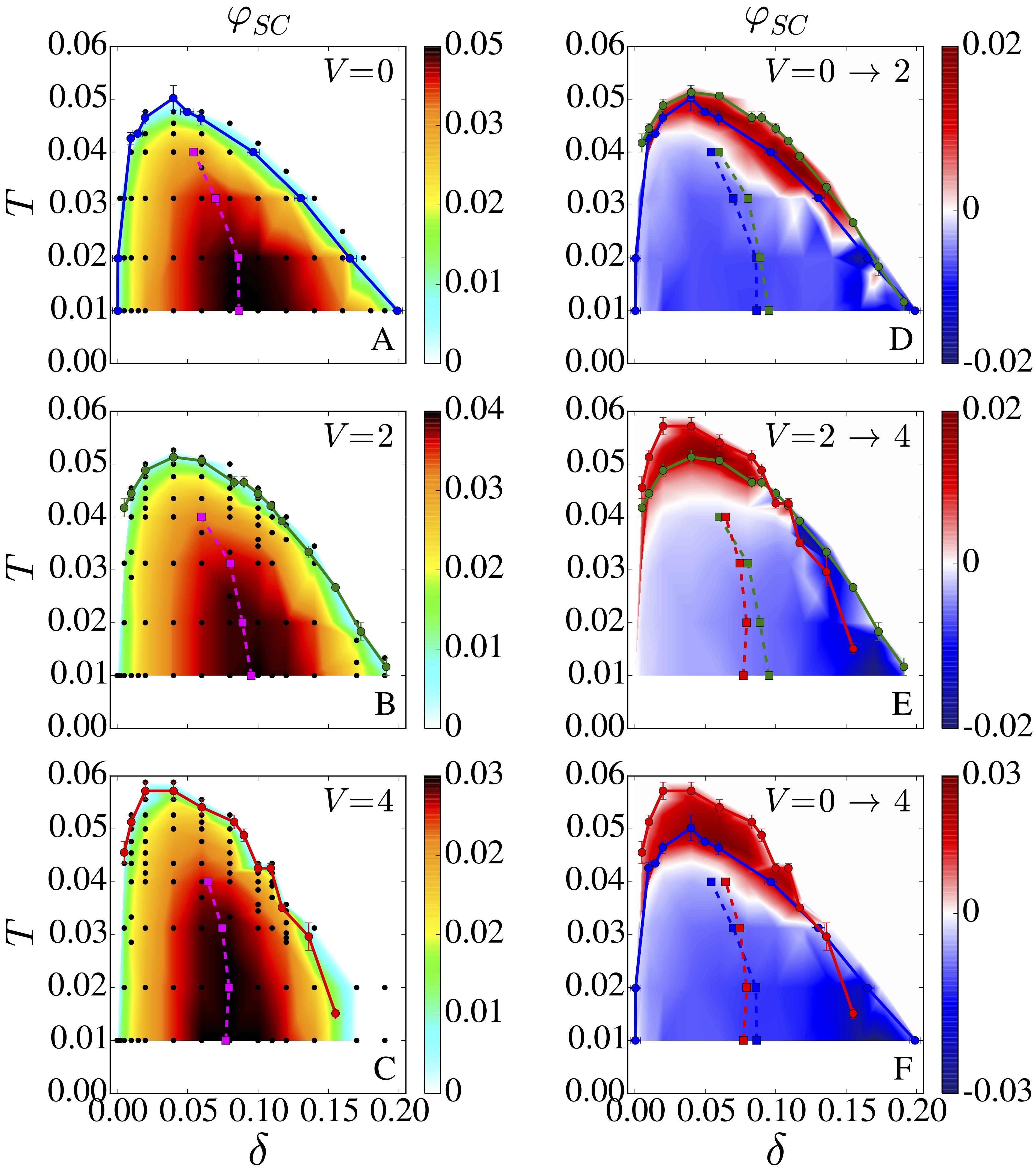}
		\caption{A, B, C: Superconducting phase diagram as a function of temperature and doping for different values of $V$. Color represents the value of the superconducting order parameter $\varphi_{SC}$ for a given doping and temperature. The black points are the data points. The color maps interpolate linearly between these points. The colored lines with dots give the transition temperature. D, E, F: Color gives the difference between $\varphi_{SC}$ at two different values of $V$. For example, the panel D gives the difference between $\varphi_{SC}$ at $V=2$ and at $V=0$. The two colored solid lines on each plot indicate the transition temperature with the same color coding as that on the panels A, B and C. The dashed lines marks the dopings where the superconducting $\varphi_{SC}$ is maximal for given temperatures.}
		\label{Fig_phi}
	\end{center}
\end{figure}

The panels A, B and C of Fig.~\ref{Fig_phi} display, on a color scale, the values of the superconducting order parameter within the superconducting domes for three values of $V$. The results for $V=0$ have appeared in Ref.~\onlinecite{Fratino:2016} but they are reproduced here for convenience. The critical temperatures bounding these domes are obtained by assuming that if the superconducting order parameter $\varphi_{SC}$ is less than $10^{-3}$, then we are in the normal state. Clearly, the critical temperature $T_c^d$ is not proportional to the low-temperature value of the superconducting order parameter. A more in-depth study of this question and comparisons with BCS theory is the subject of future work. 

To highlight the effect of $V$, the panels D, E and F of Fig.~\ref{Fig_phi} show for each temperature and doping the difference between the values of the order parameter $\varphi_{SC}$ for different values of $V$. For example, the panel D presents the difference between the values of $\varphi_{SC}$ at $V=2$ and at $V=0$. The blue regions show that at low temperature, $\varphi_{SC}$ is lower for a larger $V$ at any doping. However, $\varphi_{SC}$ becomes greater than its value at lower $V$ at high temperature as one approaches half-filling, as shown by the red regions. So $V$ enhances the resilience of superconductivity to temperature at low doping, even if it weakens $\varphi_{SC}$ at low temperature for all dopings. This has consequences on how $T_c^d$ depends on $V$. On the panels D, E and F, there are two $T_c^d$ lines with dots, one for each value of $V$. The color coding of these lines corresponds to the color coding of the corresponding lines on the panels A, B and C. One sees that $T_c^d$ is increased by $V$ at low doping but is decreased by $V$ at large doping. 

\subsection{A retarded pairing mechanism} \label{sec:retardation}

\begin{figure}[h!]
	\begin{center}
		\includegraphics[width=0.49\textwidth]{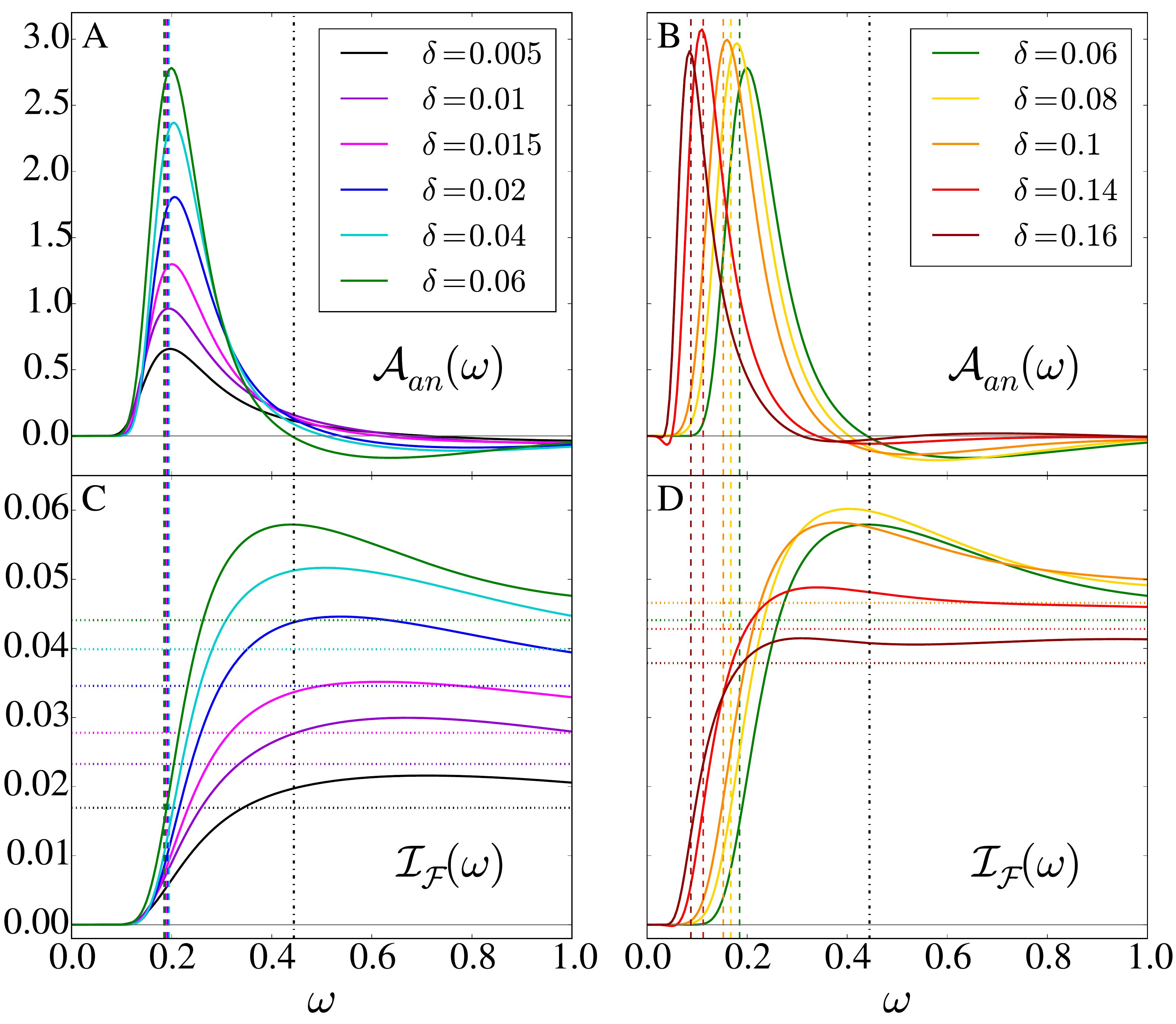}
		\caption{Hole doping evolution of the anomalous spectral function and the cumulative order parameter for $\beta = 100$ and $V = 0$. The positive-frequency part of the anomalous spectral weight is in the panels A and B: A for the underdoped regime and B for the overdoped regime. The vertical dashed color lines indicate the superconducting gap $\Delta_{SC}$ extracted from the local density of states, while the black dot-dashed lines are at $J = 4t^2/U$. The panels C and D display the cumulative order parameter obtained from the integral of the corresponding anomalous spectral weight on top. The horizontal dotted color lines in the panels C and D indicate the values taken by the superconducting order parameter $\varphi_{SC}$.}
		\label{Fig_ASF}
	\end{center}
\end{figure}

To understand the above results more deeply, consider the dynamics revealed by the anomalous spectral function. The panels A and B of Fig.~\ref{Fig_ASF} show typical anomalous spectral functions for positive frequencies at $\beta=100$ and $V=0$. The underdoped regime is on the panel A and the overdoped one on the panel B. Similar structures are found for $V=2$ and $V=4$ (not shown). A small gap is present at very low frequency, especially in the underdoped regime. It originates from short-ranged spin-singlet order that survives in our model where there is no magnetic frustration.~\cite{Sordi:2012, Merino:2014} We verified that this gap disappears in the presence of frustration.\footnote{These results were obtained from calculations performed by Charles-David H\'ebert\protect\cite{CDH:2015} for the anisotropic triangular lattice at $t'=0.4t$ and $t'=0.8t$.}  Beyond this gap, the anomalous spectral function is positive and peaks at low frequency, changes sign at a sign-changing frequency $\omega_{SC}^{\mathrm{sign}}$, and finally approaches zero at high frequency. This sign change is important. A similar sign change is observed in the Eliashberg-McMillan phononic pairing glue, attractive at low frequency and repulsive at high frequency. In the same way, the positive (negative) part of the anomalous spectral function spreads across energies where pairing (depairing) occurs.

\begin{figure}[h!]
	\begin{center}
		\includegraphics[width=0.49\textwidth]{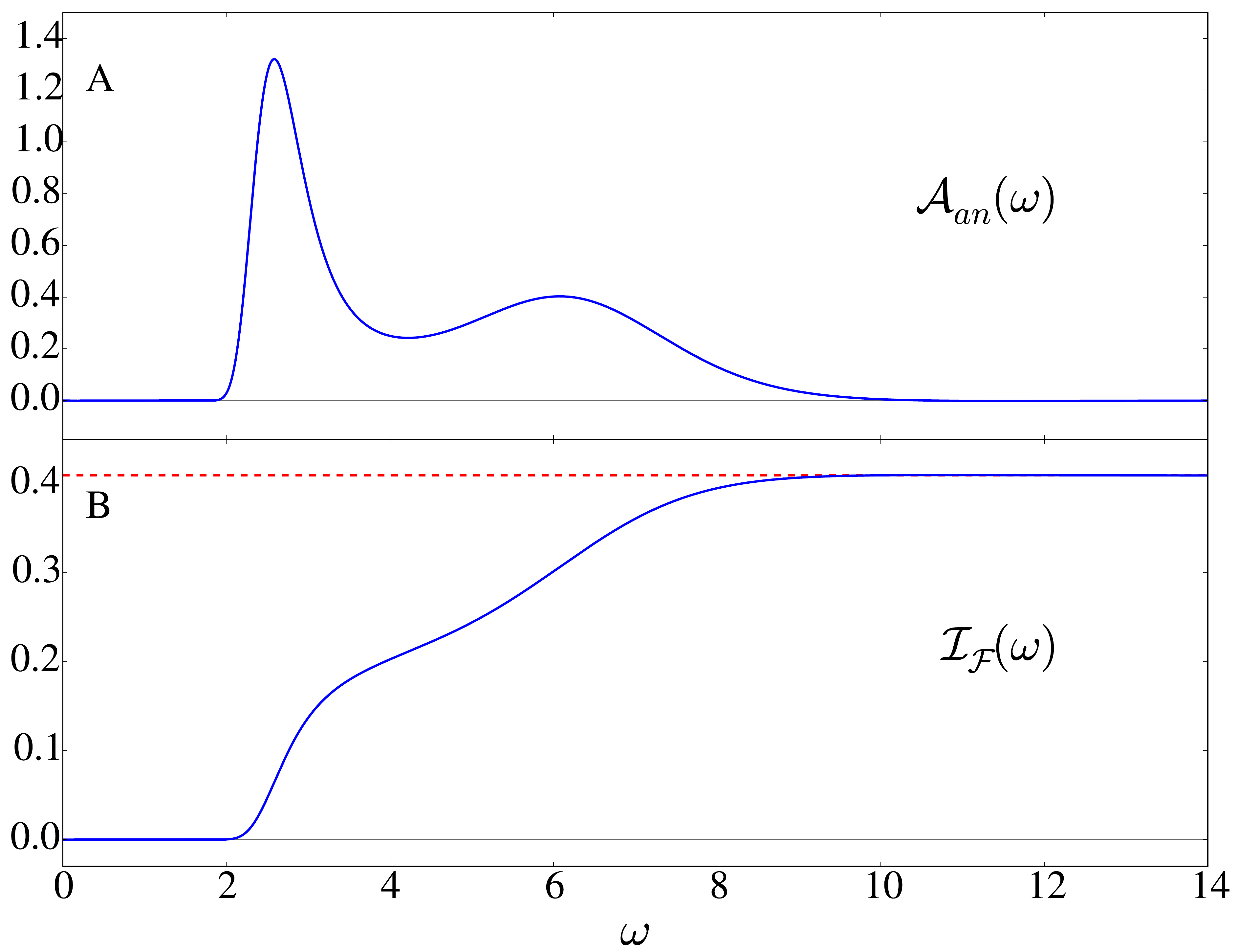}
		\caption{Pairing dynamics of the attractive Hubbard model for $U=-9$, $V=0$ and $\beta=100$. Panel A: anomalous spectral function. Panel B: cumulative order parameter. Red dashed line: asymptotic value of the cumulative order parameter.}
		\label{Fig_Hubbard_attractive}
	\end{center}
\end{figure}

It is apparent from Fig.~\ref{Fig_ASF} that pairing occurs over a small frequency range compared with the bandwidth $W \equiv 8$, which indicates a retarded pairing mechanism. Indeed, an instantaneous pairing mechanism would present pairing contributions extending over the whole bandwidth. This has been demonstrated with CDMFT simulations (solved with exact diagonalization) of the attractive Hubbard model on a two-dimensional square lattice at zero temperature,~\cite{Kyung:2007_unpublished, Kyung:2009} where $s$-wave pairing can be instantaneous. The cumulative order parameter increases over the whole bandwidth since without electronic repulsion there are no pair-breaking processes. The finite temperature pairing dynamics of this model at $U=-9t$ and 4\% doping is shown in Fig.~\ref{Fig_Hubbard_attractive} and is in complete agreement with the zero-temperature results.

Going back to Fig.~\ref{Fig_ASF}, although analytic continuation is less reliable at high frequency, the fact that the high-frequency cumulative order parameter, shown on the panels C and D of Fig.~\ref{Fig_ASF}, recovers the value of the superconducting order parameter for all $V$, independently calculated via
\begin{equation}
\varphi_{SC} = -\frac{2}{\beta} \sum_{n =0}^{+\infty} \mathrm{Re}\, \mathcal{F}(\vec{k}=(\pi,0),i\omega_n) \geq 0 \, ,
\label{Eq_order_parameter}
\end{equation}
tells us that additional high-frequency structures are unlikely. The vertical dashed color lines in Fig.~\ref{Fig_ASF} indicate the superconducting gap $\Delta_{SC}$. This gap $\Delta_{SC}$, independently extracted from half of the distance between the coherence peaks of the local density of states, generally coincides with the frequency giving the maximum of the anomalous spectral function. From this perspective, the superconducting gap is the energy where pairing is maximum. Note that $\Delta_{SC}$ saturates in the low doping regime (panels A and C of Fig.~\ref{Fig_ASF}) and decreases as hole doping increases in the large doping regime (panels B and D of Fig.~\ref{Fig_ASF}). 

\subsection{A magnetic pairing mechanism}\label{sec:magnetic}

How can $V$ both favor and disfavor superconductivity? When pairing is mediated by the Heisenberg exchange $J$, a physically reasonable explanation is~\cite{SenechalResilience:2013} that while $V$ strengthens pair-breaking Coulomb repulsion at all dopings, it also strengthens the pair-forming exchange interaction since, in the presence of $V$, the effective exchange is 
\begin{equation}
J = \frac{4t^2}{U-V} \, .
\label{Eq_J}
\end{equation}
If the retarded pairing mechanism shown in Fig.~\ref{Fig_ASF} originates from spin-fluctuation exchange, $V$ should reinforce superconductivity at low doping (where the effect of $J$ is most important) since $J$ increases with $V$, while it should weaken superconductivity at large doping, where only the pair-breaking effect of the strengthened Coulomb repulsion remains. The crucial role of $J$ in the pairing dynamics within CDMFT has been documented before.~\cite{Kancharla:2008, Civelli:2009} 

\begin{figure}[t!]
	\begin{center}
		\includegraphics[width=0.49\textwidth]{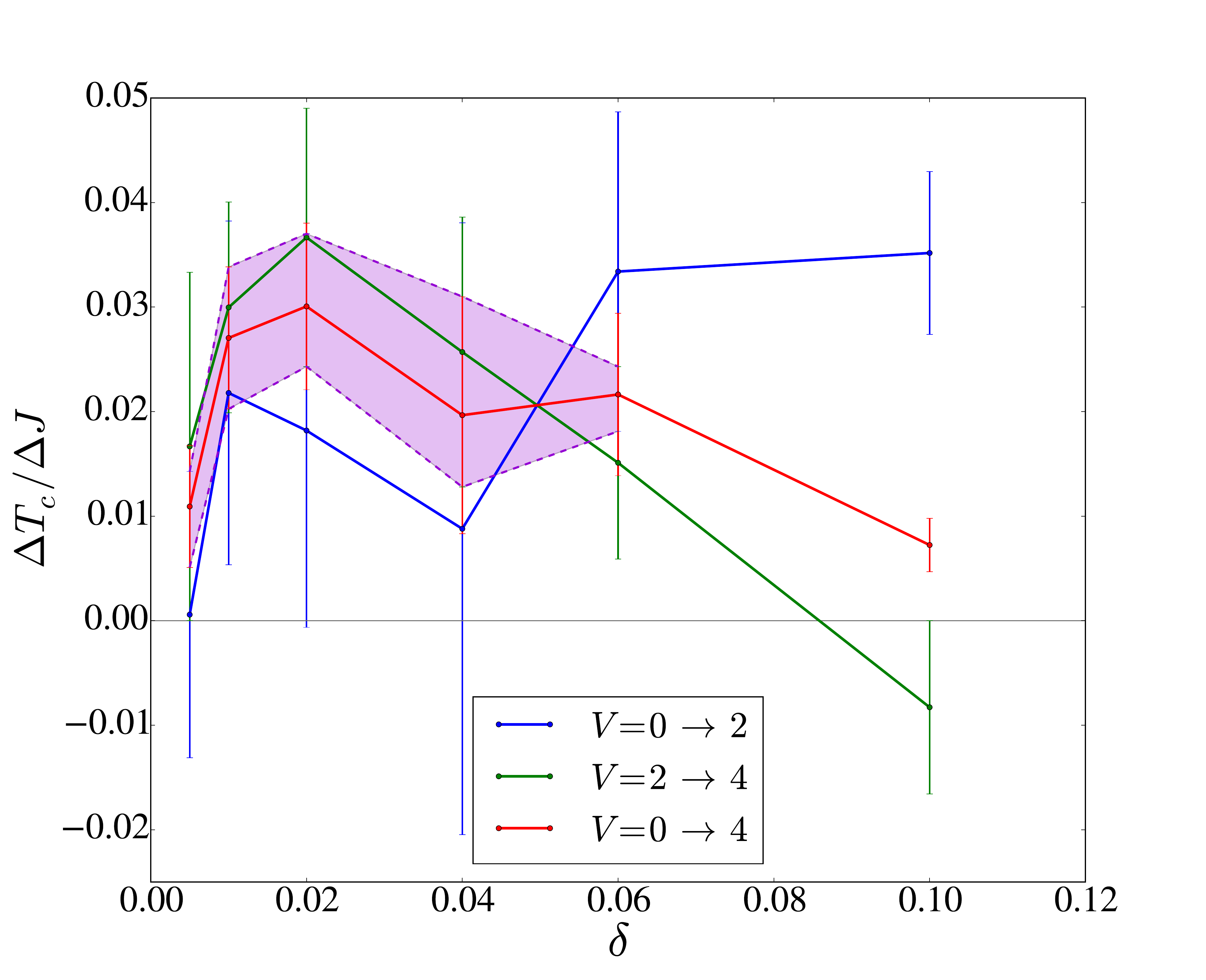}
		\caption{Difference in critical temperature divided by the difference in $J$ resulting from a change in $V$. Three cases, $V = 0$ to $2$, $V = 2$ to $4$, and $V = 0$ to $4$, are shown. While the error bars are quite large (coming from the error bars on the critical temperatures), the dark violet area corresponds to the zone where the three cases match within these error bars.}
		\label{Fig_Scaling_Tc_J}
	\end{center}
\end{figure}

Another manifestation of such a relationship between $J$ and pairing would naively appear in a scaling of the critical temperature $T_c$ with $J$. Fig.~\ref{Fig_Scaling_Tc_J} displays the difference in critical temperature divided by the difference in $J$ resulting from a change in $V$. In the case of a perfect scaling of $T_c$ with $J$, this ratio should remain constant. To help visualize our results, the dark violet area in Fig.~\ref{Fig_Scaling_Tc_J} corresponds to the zone where the values of the three studied ratios match within the error bars. In the low doping regime, where $J$ is relevant, we find that this matching area is of a reasonable size compared to the error bars, which  points toward a scaling of $T_c$ with $J$. Notice that this scaling should obviously be doping-dependent. Even in a simple BCS picture, $T_c$ would surely not scale directly with $J$ since there would be an exponential factor depending on doping through the electronic density of states. Besides, we must emphasize that spin fluctuations are more itinerant at $U=9$ than at a larger value of $U$, where the $t$-$J$ model is more relevant. This could prevent a perfect scaling of $T_c$ with $J$ at $U=9$. 
Finally, at large doping in Fig.~\ref{Fig_Scaling_Tc_J}, the scaling of $T_c$ with $J$ cannot be found anymore as $J$ becomes less relevant.

To reinforce our interpretation that $J$ drives the retarded pairing mechanism shown in Fig.~\ref{Fig_ASF}, a strong correlation between the position of the peaks in the imaginary part of the anomalous self-energy (pairing dynamics) and the position of the peaks in the imaginary part of the local spin susceptibility (spin dynamics) has also been demonstrated at low energy when $T=0$.\cite{Kyung:2009} Furthermore, a simple comparison of the anomalous spectral functions that we obtain with a) the imaginary part of the local spin susceptibility for different dopings in Ref.~\onlinecite{Kyung:2009} and b) the imaginary part of antiferromagnetic spin susceptibility at low doping for different values of $V$ in Ref.~\onlinecite{SenechalResilience:2013}, leads to deeper insight into the meaning of the characteristic frequencies of the pairing dynamics. Indeed, the frequency where the anomalous spectral function peaks has the same doping and $V$ dependence and is of the same order of magnitude as the frequency of the dominant low-frequency peak in the spin susceptibilities. Similarly, our sign-change frequency $\omega_{SC}^{\mathrm{sign}}$ is, for all dopings, roughly equal to the frequency where the dominant low-frequency peak in the spin susceptibilities ends.  That frequency is mostly $V$-independent. All of this can be understood if pairing is mediated by spin-fluctuation exchange: indeed the strength of pairing processes should come from the dominant low-frequency peak in the spin susceptibility since that peak disappears in the normal state, as shown in Ref.~\onlinecite{Kyung:2009}. Here, we study in more details where pair-breaking and pair-forming effects dominate in the phase diagram.

To disentangle the ranges of frequencies that enhance the superconducting order parameter from those that reduce it, we refer to the definition of the cumulative order parameter Eq.~\eqref{I_F} and its value at infinite frequency Eq.~\eqref{I_F_inf}. Noticing that the anomalous spectral weight changes sign at a single frequency $\omega_{SC}^{\mathrm{sign}}$, we define $\mathcal{C}_{SC}^+$
\begin{equation}
\mathcal{C}_{SC}^+ = \int_{0}^{\omega_{SC}^{\mathrm{sign}}} \! \frac{\mathrm{d}\omega}{2\pi} \, \mathcal{A}_{an}(\omega)\, f(-\omega) \geq 0
\label{Eq_Pos_cont}
\end{equation}
as the positive contribution to the order parameter and
\begin{equation}
\mathcal{C}_{SC}^- =  \left\vert\int_{\omega_{SC}^{\mathrm{sign}}}^{+\infty} \! \frac{\mathrm{d}\omega}{2\pi} \, \mathcal{A}_{an}(\omega)\, f(-\omega) \right\vert\, .
\label{Eq_Neg_cont}
\end{equation}
as the magnitude of the negative contribution. $\mathcal{C}_{SC}^-$ would be negative if it were not for the absolute value. Within the studied temperature range, the Fermi distribution in Eq.~\eqref{I_F} gives little weight to the negative frequency range of the cumulative order parameter so we do not need to integrate from $-\infty$. Physically, the fact that
	\begin{equation}
	\mathcal{C}_{SC}^+ - \mathcal{C}_{SC}^- = \varphi_{SC}
	\label{C+_C-_Interpretation}
	\end{equation}
justifies the interpretation of $\mathcal{C}_{SC}^+$ ($\mathcal{C}_{SC}^-$) as the effective strength of pair-forming (pair-breaking) processes.

\begin{figure}[t!]
	\begin{center}
		\includegraphics[width=0.49\textwidth]{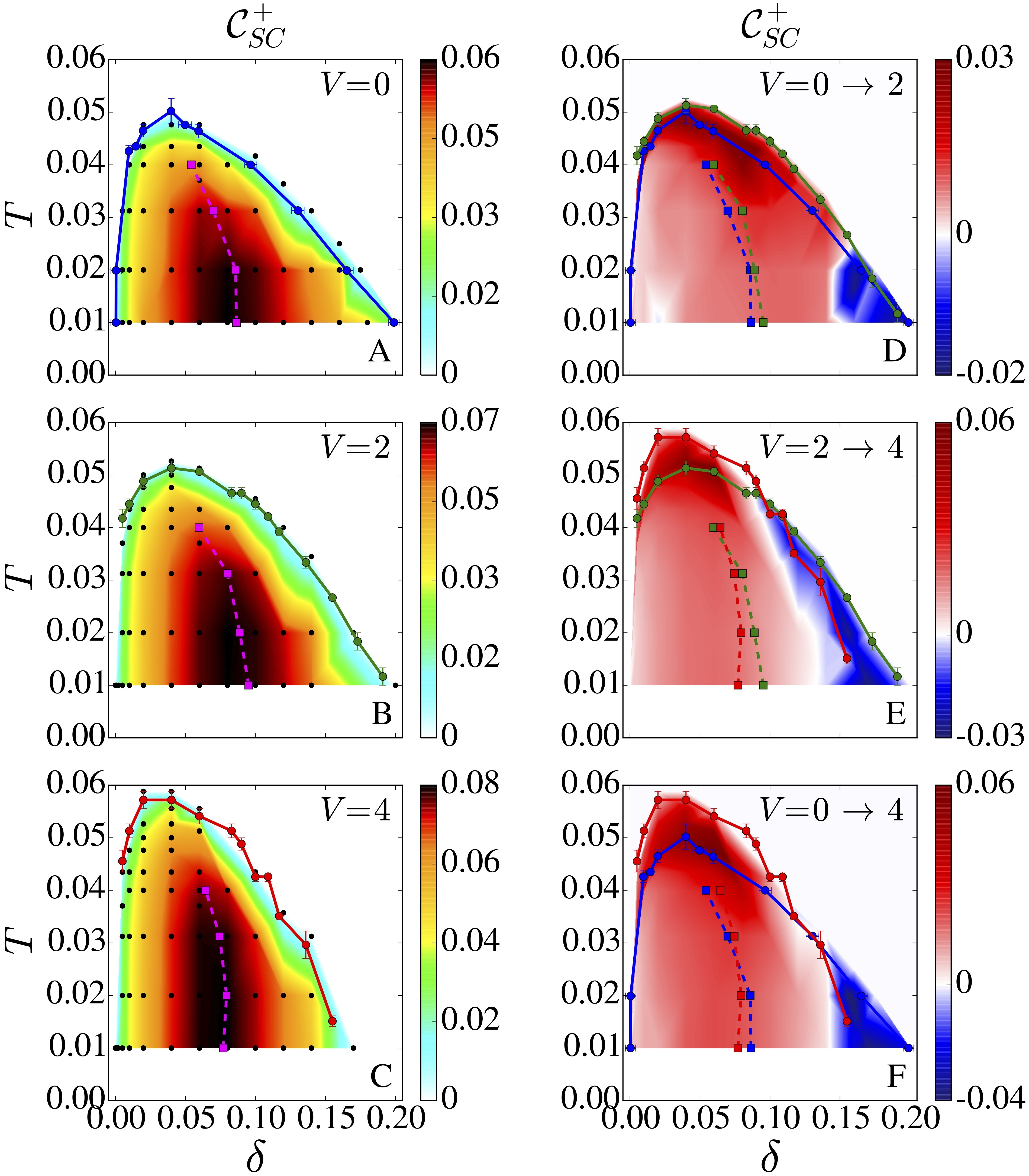}
		\caption{Analog to Fig.~\ref{Fig_phi} but where the color maps refer to the strength of pair-forming processes $\mathcal{C}_{SC}^+$ Eq.~\eqref{Eq_Pos_cont} instead of the superconducting order parameter. Solid lines: critical temperatures for given dopings. Dashed lines: dopings where the superconducting order parameter is maximal for given temperatures.}
		\label{Fig_pos_cont}
	\end{center}
\end{figure}

\begin{figure}[t!]
	\begin{center}
		\includegraphics[width=0.49\textwidth]{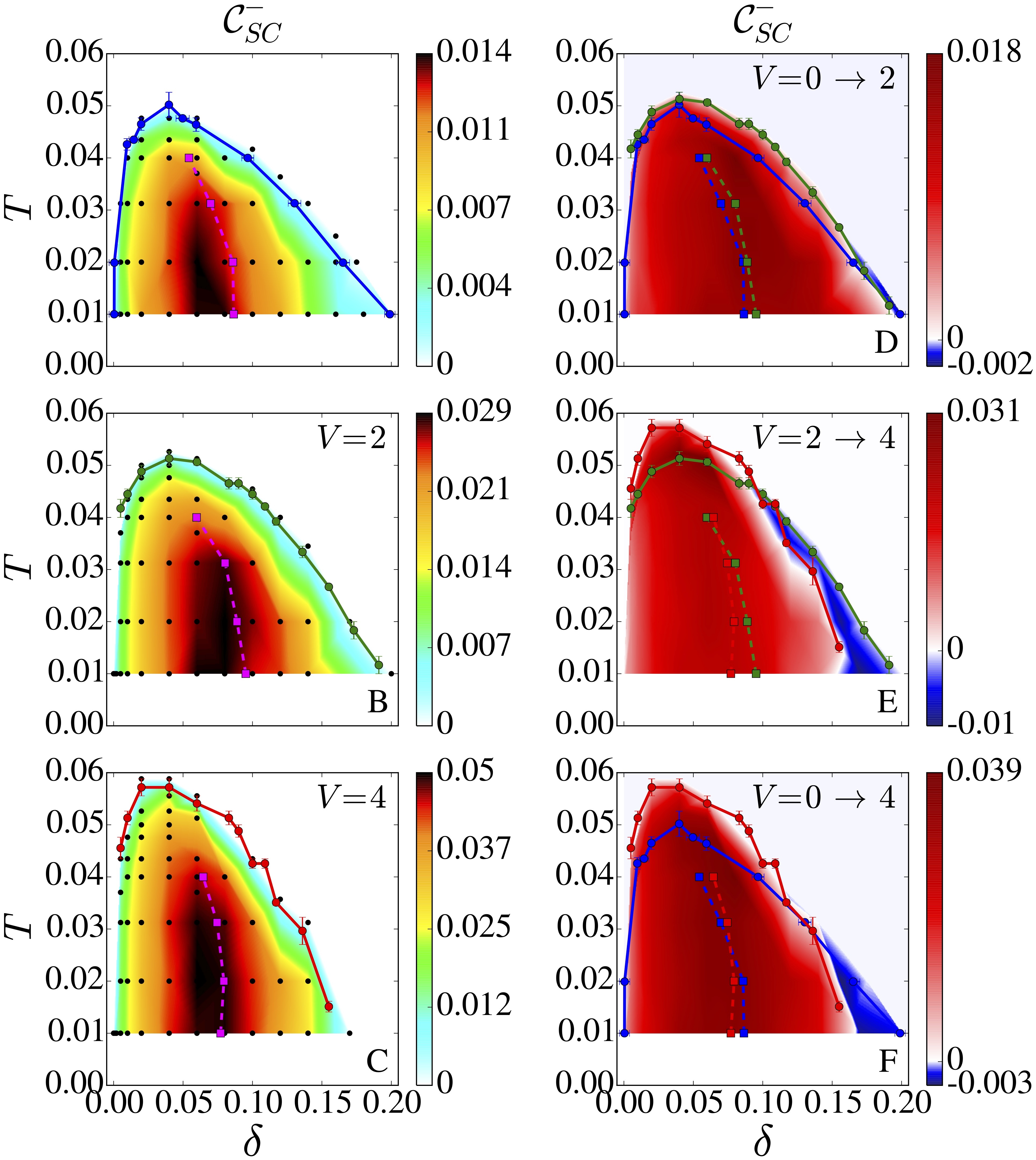}
		\caption{Analog to Fig.~\ref{Fig_phi} but where the color maps refer to the strength of pair-breaking processes $\mathcal{C}_{SC}^-$ Eq.~\eqref{Eq_Neg_cont} instead of the superconducting order parameter. Solid lines: critical temperatures for given dopings. Dashed lines: dopings where the superconducting order parameter is maximal for given temperatures.}
		\label{Fig_neg_cont}
	\end{center}
\end{figure}

Fig.~\ref{Fig_pos_cont} and Fig.~\ref{Fig_neg_cont} present, respectively, the values of the positive and negative contributions to the order parameter and their variations with $V$ within the superconducting domes. The panels D, E and F of these figures illustrate our previous discussion. On the one hand, the positive contribution $\mathcal{C}_{SC}^+$ always increases with $V$ at low doping, reflecting the beneficial effect on superconductivity induced by the strengthening of nearest-neighbor $J$, but drops with $V$ at large doping, where $J$ is less relevant. On the other hand, the magnitude of the negative contribution $\mathcal{C}_{SC}^-$ always increases with $V$ for any doping (the blue areas where $\mathcal{C}_{SC}^-$ apparently decreases with increasing $V$ can be misleading since they come from the difference in $T_c^d$). This is expected since the detrimental effect of pair-breaking Coulomb repulsion on superconductivity is always strengthened by $V$, at least in the simple interpretation Eq.~(\ref{C+_C-_Interpretation}) given above. More discussion on this may be found in Sec.~\ref{sec:discussion}\textcolor{Red}{.}\ref{sec:pairing}. 

%

\subsection{A quantitative view of instantaneous pair-breaking processes} \label{sec:instantaneous}

We have seen that one could extract a dynamical negative contribution to pairing for frequencies larger than the frequency $\omega_{SC}^{\mathrm{sign}}$ where the anomalous spectral function changes sign. However, the instantaneous contribution to pairing does not come out clearly from this analysis. Indeed, the results from the attractive Hubbard model presented in Fig.~\ref{Fig_Hubbard_attractive} confirm that this contribution would show up through certain pairing contributions extending over the whole bandwidth.

\begin{figure}[h!]
	\begin{center}
		\includegraphics[width=0.49\textwidth]{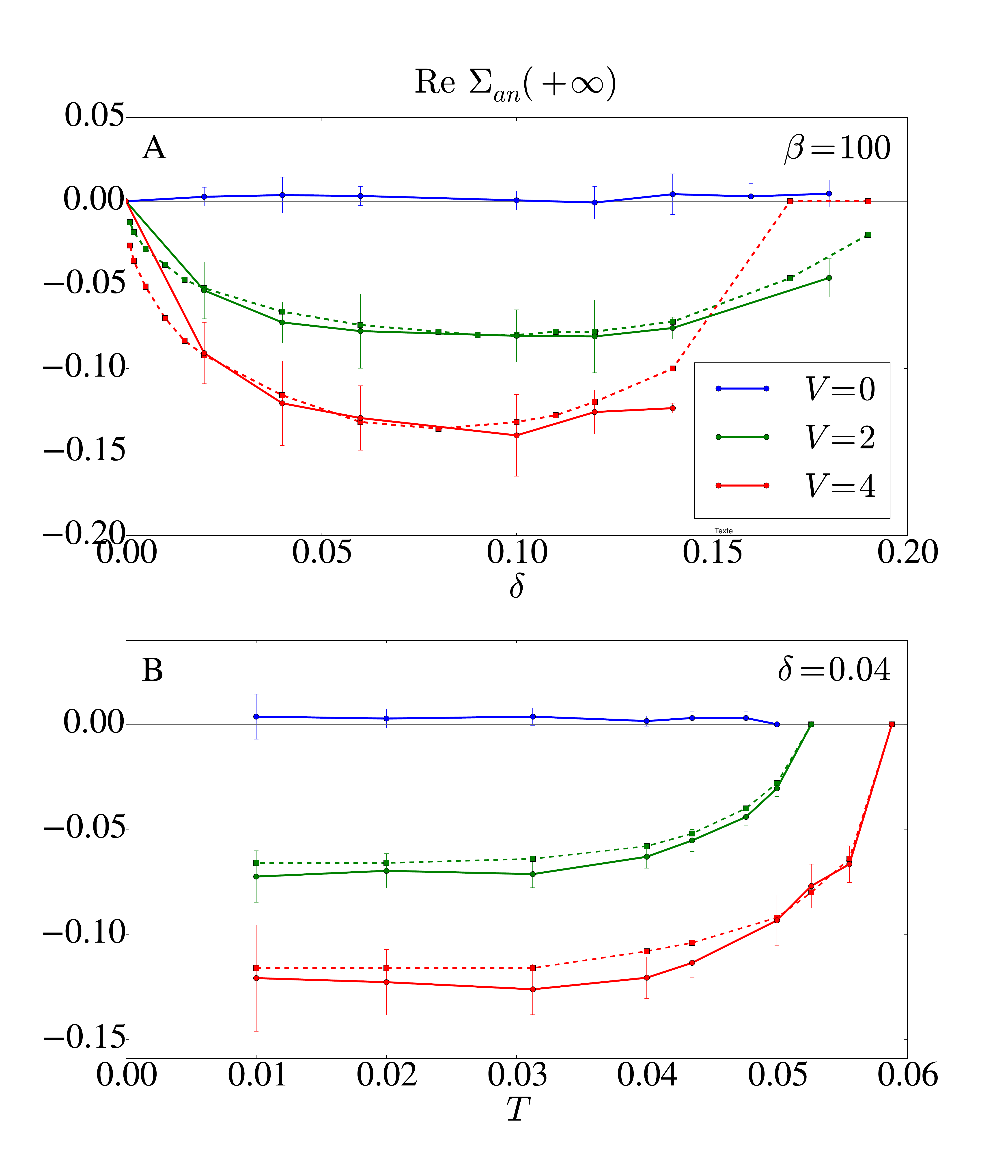}
		\caption{Infinite-frequency contribution to the real part of the anomalous self-energy for different values of $V$ (the color code is the same than the one used previously). A: As a function of hole doping at $\beta=100$. B: As a function of temperature at 4\% hole doping. Dashed lines: anomalous Fock contributions Eq.~\eqref{Eq_Fock} for given values of $V$.}
		\label{Fig_cont_self}
	\end{center}
\end{figure}

A clearer way to extract this information is through the anomalous self-energy $\Sigma_{an}(i\omega_n) \equiv \Sigma_{an}(\vec{k}=(\pi,0), i\omega_n)$ defined via the inverse of the Green's function matrix in Nambu formalism
\begin{eqnarray}
 & & \mathrm{G}^{-1}(\vec{k},i\omega_n) \nonumber \\
 & = & 
\begin{pmatrix}
i\omega_n - \xi_{\vec{k}} - \Sigma(\vec{k}, i\omega_n) & -\Sigma_{an}(\vec{k},i\omega_n) \\
-\Sigma_{an}(\vec{k},i\omega_n) & i\omega_n + \xi_{-\vec{k}} + \Sigma(-\vec{k}, -i\omega_n)
\end{pmatrix}
 \, , \nonumber \\
\end{eqnarray}
where $\xi_{\vec{k}}$ is the free dispersion relative to the chemical potential. Unlike Green's functions or hybridization functions, the anomalous self-energy is not constrained by any sum rule to vanish at high frequency, so that one writes
\begin{equation}
\Sigma_{an}(i\omega_n) = \Sigma_{an}(+\infty) + \int\! \frac{\mathrm{d}\omega}{2\pi}\, \frac{\mathrm{Im}\, \Sigma_{an}(\omega)}{i\omega_n - \omega}\, .
\end{equation}
Taking the real part of this equation gives
\begin{equation}
\mathrm{Re}\,\Sigma_{an}(i\omega_n) = \mathrm{Re}\,\Sigma_{an}(+\infty) - \int\! \frac{\mathrm{d}\omega}{2\pi}\, \frac{\omega\,\mathrm{Im}\, \Sigma_{an}(\omega)}{\omega_n^2 + \omega^2} \, .
\label{Contribution_self_an_infinite}
\end{equation}
The Matsubara-frequency-dependent anomalous self-energy is sufficient here to extract the infinite-frequency contribution $\mathrm{Re}\,\Sigma_{an}(+\infty)$ since it is identical in Matsubara and real frequency formalisms. Given that we consider a positive superconducting order parameter as sign convention here, a positive infinite-frequency contribution is favorable to superconductivity whereas a negative one is detrimental to superconductivity. The value of $\mathrm{Re}\,\Sigma_{an}(+\infty)$ is presented in Fig.~\ref{Fig_cont_self} for $\beta=100$ as a function of doping and for 4\% hole doping (optimal doping) as a function of temperature.


The infinite-frequency contribution for the $(\pi,0)$ anomalous self-energy is finite only for a finite value of $V$ and is always negative, therefore always detrimental to superconductivity. The dashed lines in Fig.~\ref{Fig_cont_self} show that this contribution is merely the anomalous Fock contribution
\begin{equation}
\mathrm{Re}\,\Sigma_{an}(+\infty) = -V\varphi_{SC}\, ,
\label{Eq_Fock}
\end{equation}
where the order parameter $\varphi_{SC}$ is defined in Eq.~\eqref{Eq_order_parameter}. Within the attractive Hubbard model,~\cite{Kyung:2007_unpublished} the infinite-frequency contribution is always positive since the superconducting processes are always pair-forming for that model.

\subsection{A system pushed towards charge ordering}\label{sec:charge}
\label{SubSec_CDW}

Even at half-filling, a sufficiently large $V$ will induce a charge-density wave since a configuration with doubly-occupied sites with no charge on their nearest-neighbors become less costly in potential energy. Indeed, at large $U$ and $V$, simple potential-energy minimization arguments show that there will be a transition to the charge-density-wave state around $U=zV$ where $z$ is the number of nearest neighbors. This has been discussed first in one dimension,~\cite{EmeryVoneD:1979} and then on the two-dimensional square lattice.~\cite{Callaway:1989}  Recent studies for small values of $U$ and $V$ show that even at finite doping, sufficiently large $V$ promotes charge ordering.~\cite{Raghu:2012} On our cluster, a large value of $V$ (with $V<U/2$) also promotes a commensurate charge order whose pattern consists in a simple alternation of empty and doubly-occupied sites. However, no symmetry other than that associated to $d$-wave superconductivity is allowed to be explicitly broken within the CDMFT bath so this order is not observed. Nevertheless, this does not mean that the system does not exhibit signatures of this tendency towards charge ordering.  

\begin{figure}[h!]
	\begin{center}
		\includegraphics[width=0.35\textwidth]{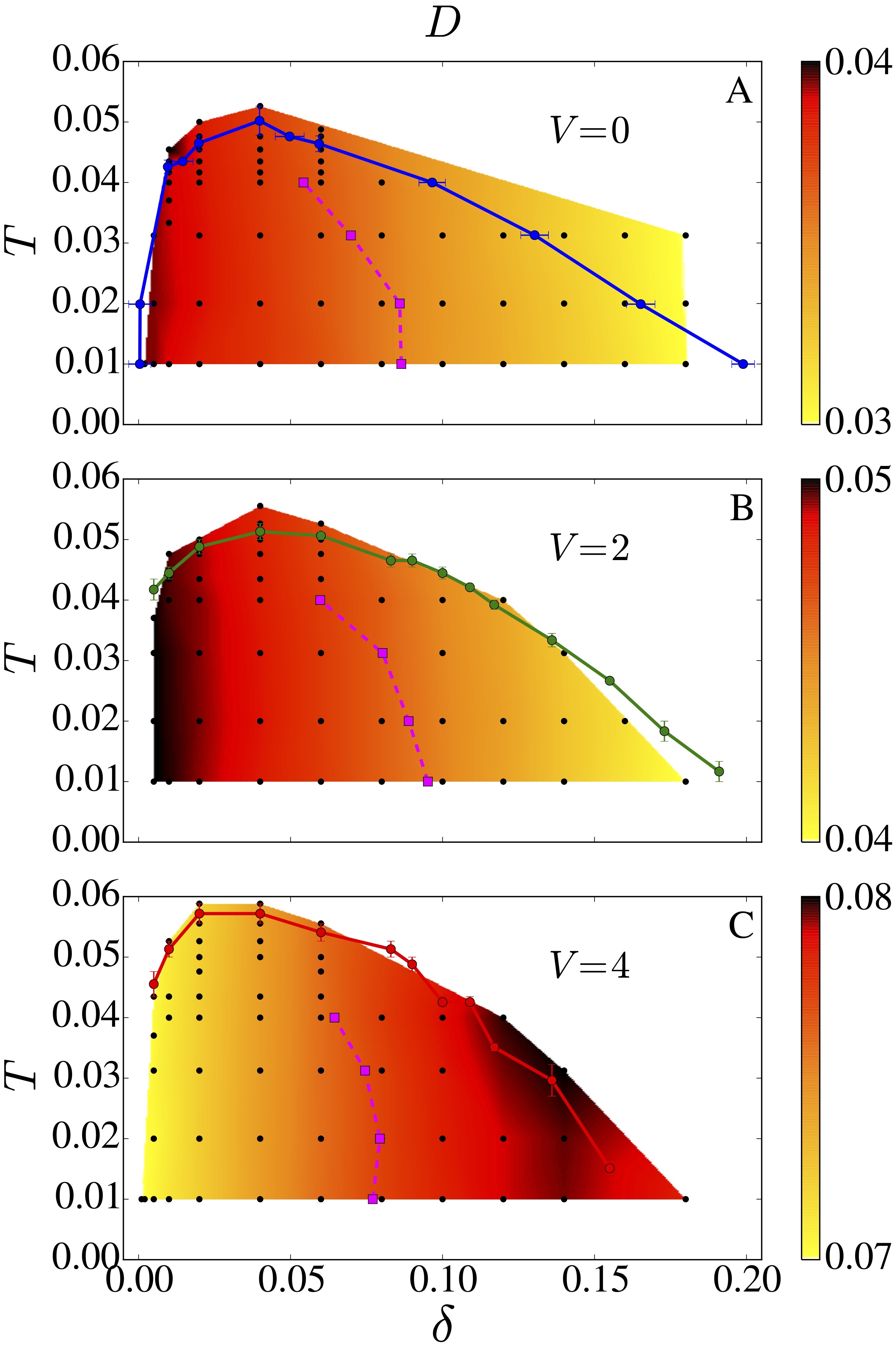}
		\caption{Analog to the panels A, B and C of Fig.~\ref{Fig_phi} but where the color maps refer to the double occupancy $D$ instead of the superconducting order parameter. Solid lines: critical temperatures for given dopings. Dashed lines: dopings where the superconducting order parameter is maximal for given temperatures. The double occupancy has been artificially put to zero outside the convex hull of the data points for the sake of clarity.}
		\label{Fig_D}
	\end{center}
\end{figure}

The tendency towards charge order manifests itself in figure Fig.~\ref{Fig_D} that presents the value of double occupancy $D = \langle \hat{n}_{i\uparrow} \, \hat{n}_{i\downarrow} \rangle_{\mathrm{cluster}}$ within the superconducting dome for different values of $V$. Qualitative trends as a function of temperature and $V$ are the same in the normal state. The double occupancy increases with $V$ over the whole doping and temperature ranges, which clearly shows that the system is pushed towards charge ordering by $V$. The doping dependence of the double occupancy for each value of $V$ gives more insight into the physics of this charge ordering. At $V=0$ and $V=2$, the double occupancy behaves as usual: as electrons are removed from the system upon doping, it becomes less and less likely to doubly-occupy a site and the double occupancy decreases when the doping increases. However, the doping dependence of $D$ is completely reversed at $V=4$. This may be understood from the competition between charge ordering and the antiferromagnetic fluctuations strengthened by $V$ through $J=4t^2/(U-V)$ at low doping. When antiferromagnetism becomes less relevant at large doping, charge should tend to order. The fact that $T_c^d$ is strongly increased at low doping for $V=4$ compared to $V=2$ might come from an increase in the tendency to pairing mediated by charge fluctuations, as seen recently in the small correlation limit.\cite{Chubukov:2015}

For extended-$s$ or $s+d$ symmetry, electrons can pair up on a single site and this on-site pairing could be enhanced by $V$ since it favors double occupancy. We verified whether the above charge fluctuations could favor these symmetries. We found that these symmetries are not stabilized by $V$ at $\beta=100$ and 4\% hole doping. Notice that the four-point updates, that have been shown essential for ergodicity,\cite{Semon:2014}  are important here not only quantitatively, as in the $d$-wave case, but also qualitatively since we found unphysical results without them. 

%
%
%

\section{Discussion} \label{sec:discussion}

\subsection{Shape of the superconducting dome without $V$}

The superconducting $T_c^d$ in plaquette CDMFT studies is asymmetrical.\cite{Fratino:2016} For 8-site clusters, $T_c(\delta)$ at $U=6t$ also shows a strong asymmetry with a maximum at rather low doping ($5\%$),\cite{GullParcolletRaman:2013} like we find. However, contrary to our results, these larger system-size studies suggest that there is a small doping range near half-filling where superconductivity disappears.\cite{gull_superconductivity_2013}

In pioneering FLEX studies,\cite{Bickers_dwave:1989} it was found that in the weak correlation limit, $T_c^d$ for $d$-wave superconductivity increases all the way to half-filling in the $t'=0$ model. However, FLEX does not lead to a pseudogap in the momentum-resolved spectral weight.\cite{Moukouri:2000} When this effect is included, then, in the $t'=0$ model, there is a superconducting dome.~\cite{Kyung:2003} In this weak correlation regime, the pseudogap is induced by long wavelength  antiferromagnetic fluctuations.\cite{Vilk:1997} In the doped-Mott insulator regime studied here, there is also a pseudogap, but it comes from short-range correlations induced by $J$. More specifically, the pseudogap appears when the plaquette singlet becomes the most probable state.\cite{Sordi:2012} This is reminiscent of RVB physics. \cite{AndersonVanilla:2004} It is this pseudogap that eventually leads to a fall of $T_c^d$. 

To understand why the fall of $T_c^d$ occurs so close to half-filling, one needs to understand where the pseudogap appears in the plaquette CDMFT $T-\delta$ phase diagram. The physics that determines where the pseudogap appears is a normal-state first-order transition that also acts as an organizing principle\cite{Fratino:2016} for the superconducting phase diagram.\footnote{This organizing principle has also been discussed in the context of layered organic superconductors.\protect{\cite{CDH:2015} }} There is a Widom line that controls crossovers at temperatures above the critical point of the first order transition.\cite{Sordi:2012} That Widom line and its precursor determine where the pseudogap appears.\cite{Sordi:2013,Alloul:2013} The Widom line is tilted towards half-filling hence the maximum $T_c^d$ is close to half-filling.  

We interpret our value of $T_c^d$ as a mean-field result that indicates where short-range pairs form and where superconducting fluctuations are important.\cite{SordiSuperconductivityPseudogap:2012} The actual $T_c$ in experiment will be influenced by Kosterlitz-Thouless physics, competing order, long-wavelength particle-hole and particle-particle fluctuations, phase fluctuations~\cite{emery_importance_1995,  hetel_quantum_2007} and disorder.\cite{Alloul:2010} For example, the fact that there is a competition between antiferromagnetism and superconductivity is clear in zero-temperature plaquette CDMFT studies.\cite{Kancharla:2008} We also note that the superconducting correlation length increases as one approaches the true $T_c$ so that finite-size effects could become more important in that regime. 

\subsection{Effect of $V$ on the superconducting dome}\label{Sec:SC_Dome}

As seen from Fig.~\ref{Fig_phi}, in the presence of nearest-neighbor repulsion $V$, we find that the doping range where superconductivity appears at low temperature is narrowed and that the maximum $T_c^d$ approaches half-filling even more. The detrimental effect of $V$ on $T_c^d$ at large doping is the expected effect coming from strengthened Coulomb repulsion. There is also a decrease of the order parameter induced by $V$ at low temperature as expected. The surprising result is that the maximum $T_c^d$ increases with $V$. This can be understood if superconductivity is controlled by $J$ since in the strong-correlation regime $V$ increases this coupling constant, $J=4t^2/(U-V)$. Even though the order parameter is decreased by $V$, in BCS theory $T_c^d$ depends on the product of the zero-temperature order parameter with the coupling constant so that an increase of $T_c^d$ is not necessarily unphysical. A smaller pseudogap with an even more tilted Widom line in the presence of $V$ could also explain the effect. All this would need further investigation.

\subsection{Pairing mechanism, retardation and glue}\label{sec:pairing}

The dynamics of pairing gives additional insight into the above results. At $V=0$, that dynamics has been extensively studied in quantum cluster methods (DCA-CDMFT) and exact diagonalization\cite{Poilblanc:2002} through the frequency-dependence of the anomalous self-energy,\cite{Maier:2008, CivelliPairingPRL:2009, CivelliPairing:2009,Kyung:2009} of the gap function\cite{Poilblanc:2002, Gull_Millis:2014} and of the anomalous spectral weight or corresponding cumulative order parameter.\cite{SenechalResilience:2013,Bzdusek:2015} The correspondence with the spectral weight of spin fluctuations\cite{Poilblanc:2002, Maier:2008, Kyung:2009, Gull_Millis:2014} gives credence to the spin-fluctuation mechanism repeatedly proposed for many years using different methods.\cite{Beal-Monod:1986, Scalapino:1986, Kotliar:1988, Monthoux:1992, Scalapino:1999, MaitiChubukovReview:2013, TremblayJulichPavarini:2013}

The effect of $V$ on the phase diagram and on the pairing dynamics confirms, for $V > 0$, the above results. Namely, the pairing dynamics is strongly retarded: in other words, pairing occurs at very low frequencies, of order $J$, and is reinforced by $V$ at low frequency for a given $U$ ($J=4t^2/(U-V)$),  while at larger frequencies $V$ plays a detrimental role, as seen from the increase with $V$ of the high-frequency negative contributions to the cumulative order parameter. The finite negative value of the anomalous self-energy at infinite frequency displayed in Fig.~\ref{Fig_cont_self} also reveals an instantaneous depairing effect of $V$ coming from anomalous Fock contributions. 

We stress that several ways have been proposed to identify pairing and non-pairing contributions to superconductivity. For example, in Eliashberg theory, the main phonon frequency~\cite{BergmannTc:1973} and the average property of the phonon spectrum~\cite{AllenDynes:1975} that most influence $T_c$ have been found. The phonon frequency that is most important for the zero temperature gap and its ratio to $T_c$ has also been found.~\cite{Mitrovic:1980} In our case, based on previous work,~\cite{Kyung:2009,SenechalResilience:2013} we chose positive and negative anomalous spectral weight to identify pairing and depairing frequency ranges instead of positive and negative contributions to the final value of $T_c$. We concluded that $V$ was depairing at high frequency. Instead, if we had the equivalent of a BCS or Eliashberg theory at our disposal with a corresponding prediction of $T_c$ from the microscopic parameters, we could have arrived at a different conclusion. For example, it has been suggested,~\cite{Akashi:2014} for an $s$-wave superconductor, that a large on-site repulsion $U$ can also increase $T_c$ in the frequency region where the gap function becomes negative since, then, the product between $U$ and the gap leads to an effective attractive interaction. Nevertheless, even if one could expect the same thing to happen with $V$, here we found that at infinite frequency, $V$ is definitely pair-breaking. Note also that since $V$ is flat in frequency it should have some pair-breaking effects at low frequency as well. With our interpretation, we found in the underdoped regime that, at low frequency, pairing effects of $V$ prevail on depairing effects.

Retardation is expected in the weak-correlation regime where the pairing mechanism is understood as arising from the exchange of antiferromagnetic fluctuations.\cite{Beal-Monod:1986,Scalapino:1986,Scalapino:1999,Scalapino_E-letter,Scalapino_RMP:2012} By contrast, in the doped Mott-insulator regime where correlations are strong, it has been suggested that there is no glue.\cite{Anderson:2007} Indeed, in that limit, the Hubbard model can be approximated by the $t$-$J$ model where $d$-wave pairing can be found in a mean-field factorization of Heisenberg exchange, suggesting instantaneous pairing~\cite{Kotliar:1988,Poilblanc:2002} by analogy with the $s$-wave case in the attractive Hubbard model. As discussed above, the latter picture is not correct. Nevertheless, even in an instantaneous pairing picture, $V$ could have both pairing effects through $J$ and direct depairing effects. While the antagonistic effects of $V$ on the pairing dynamics are not sufficient to distinguish between the retarded and instantaneous pairing pictures, the frequency dependence and pairing range are sufficient, as discussed in Fig.~\ref{Fig_ASF} and Fig.~\ref{Fig_Hubbard_attractive}.  

There are nevertheless differences in the pairing mechanism in the weak and strong correlation regime.\cite{TremblayJulichPavarini:2013} In the weak correlation limit, it is long-wavelength antiferromagnetic fluctuations that mediate pairing. In quantum-cluster studies of the strong correlation limit, spin fluctuations are short-ranged and can hardly be distinguished from spin fluctuations due to local singlets or RVB physics, even though amongst the four wave vectors of the cluster, it is $(\pi,\pi)$ that dominates. Another way to distinguish between the physics at weak and strong correlations is to identify whether the condensation energy originates from a gain in potential energy, as in BCS theory, or from a gain in kinetic energy.\cite{Anderson:1987a} This criterion does not rely on detailed dynamical considerations since kinetic energy, for example, depends on an integral over frequency of the spectral weight. Quantum cluster studies suggest that for strong correlation condensation energy comes from kinetic energy~\cite{Maier:2004,Fratino:2016} whereas at intermediate correlation strength, there can be a crossover from kinetic-energy driven to potential-energy driven as doping increases.\cite{Gull_Millis:2012,Fratino:2016} See Ref.~\onlinecite{VanDerMarelOptical:2016} for a recent discussion of the experimental situation.

\section{Conclusion} 

A finite nearest-neighbor repulsion $V$ has antagonistic effects on the plaquette CDMFT phase diagram for $d$-wave superconductivity in the doped Mott-insulator regime ($U=9t$). In the zero-temperature limit, $V$ decreases the superconducting order parameter more and more with over-doping.\cite{SenechalResilience:2013} Our finite temperature studies have allowed us to show that although $V$ decreases $T_c^d$ in the overdoped regime, as expected, it increases $T_c^d$ in the underdoped regime. This cannot be excluded on physical grounds since, even in BCS theory, $T_c^d$ depends on the product of the order parameter and of the pairing strength. Hence, a decrease of the order parameter concomitant with an increase in $T_c^d$ simply reflects the increase in pairing strength expected as we approach half-filling when the pairing mechanism involves superexchange $J$.  A clue towards understanding how $V$ can favor pairing through $J$, is that $J=4t^2/(U-V)$. The decrease in $T_c^d$ at large doping is the behavior expected from strengthened Coulomb repulsion. 

Our investigation of the frequency-dependent anomalous spectral function at finite-temperature has allowed us to further disentangle the paradoxical role of $V$. This spectral function exhibits a positive part and a negative part, respectively associated with pair-forming and pair-breaking physical processes, from which we extract positive and negative contributions to the pairing dynamics. The positive contribution $\mathcal{C}_{SC}^+$ increases with $V$ at low doping but decreases with $V$ at large doping, whereas the negative contribution $\mathcal{C}_{SC}^-$ increases with $V$ at any doping. While the negative contribution is easily explained by the systematic strengthening of pair-breaking Coulomb repulsion $V$, the positive contribution comes out of low-frequency pair-forming fluctuations induced by the coupling constant Eq.~\eqref{Eq_J}. However, the latter beneficial effect on superconductivity is less relevant at large doping where $J$ becomes less important. Hence, the clue towards resolving the antagonistic effects of $V$ on pairing resides in the retarded nature of the pairing interaction. Indeed, the energy where the cumulative order parameter is maximum occurs at a small value (of order $J$) compared to the bandwidth while the pair-breaking effect of $V$ occurs at larger energy scales, in agreement with zero-temperature results.\cite{SenechalResilience:2013} In addition, there is an instantaneous pair-breaking contribution coming from $V$, as demonstrated by the value of the anomalous self-energy at infinite-frequency. Retardation is thus crucial not only for weak correlations\cite{SenechalResilience:2013} but also for strong correlations. We also found signs that charge order should become important at large $V$ and that $d$-wave superconductivity is always preferred over extended-$s$ and $s+d$-wave. 

Future studies should include the competing effects of antiferromagnetism at finite temperature and improved approximations for the inter-cluster effects of $V$. It would also be important to include magnetic frustration through next-nearest-neighbor hopping amplitude, although this may worsen the sign problem in CTQMC.  

\section*{Acknowledgments}

We are indebted to S. Verret, D. S\'en\'echal, J. Gukelberger, D. Bergeron and R. Nourafkan for fruitful discussions. This work has been supported by the Natural Sciences and Engineering Research Council of Canada (NSERC) under grant RGPIN-2014-04584, and by the Tier I Canada Research Chair Program (A.-M.S.T.). Simulations were performed on computers provided by CFI, MELS, Calcul Qu\'ebec and Compute Canada.


%

\end{document}